\documentclass[]{tsfp6}
\usepackage{verbatim} 

\title{Intermittency effects in rotating decaying turbulence}

\author{{\bfseries Arkadiusz K. Kuczaj}\\
Nuclear Research and Consultancy Group (NRG)\\
P.O. Box 25, 1755 ZG Petten, the~Netherlands\\
arek@kuczaj.pl\\
\vspace{24pt}
{\bfseries Bernard J. Geurts}\\
Multiscale Modeling and Simulation, NACM, J.M. Burgers Center, University of Twente \\
P.O. Box 217, 7500 AE Enschede, the Netherlands \\ Fluid Dynamics Laboratory, Department of Applied Physics, Eindhoven University of Technology \\
P.O. Box 513, 5300 MB Eindhoven, the~Netherlands \\
\vspace{24pt}
{\bfseries Darryl D. Holm}\\
Mathematics Department, Imperial College London \\
SW7 2AZ, London, UK \\
CCS2, Los Alamos National Laboratory \\
Los Alamos, NM 87545, USA
}

\begin{document}

\maketitle

\thispagestyle{empty}
\section{abstract}
Rotation modulates turbulence causing columnar structuring of a turbulent flow in case of sufficiently strong rotation. This yields significant changes in the flow characteristics and dispersion properties, which makes rotational turbulence modulation particularly relevant in the context of atmospheric and oceanic flows. Here we investigate the canonical flow of turbulence in a periodic box, subjected to rotation about a fixed vertical axis.  As point of reference we consider direct numerical simulations of homogeneous isotropic turbulence. Modulation due to rotation at various rotation rates (i.e., different Rossby numbers) is investigated. Special attention is paid to the alteration of intermittency, which is measured in terms of changes in the scaling of the structure functions. A reduction of intermittency quantified with the longitudinal structure functions in the direction perpendicular to the rotation axes will be presented. These numerical findings correspond well to recent results obtained in experiments by Seiwert et al. (2008)~\cite{seiwert:decrease:2008}. 

\section{Turbulence and rotation}
Turbulence exhibits intense bursts of vorticity and strain that  can be important for example in production of forceful vortices in atmospheric flows. An important interest in atmospheric fluid dynamics is concentrated around  the impact of the Coriolis force on turbulence that tends to  the two-dimensionalization of the flow. In this paper we consider direct numerical simulations of decaying turbulence in a rotating frame of reference studying the effect of the Coriolis force on turbulence. For high rotation rates the Coriolis force is dominant in a wide range of scales and plays an important role balancing the convective nonlinearity and viscous forces~\cite{chen:resonant:2005}. We will show that the Coriolis force not only suppresses the forward energy transfer to small scales, but also  modifies the dynamics of turbulence measured in terms of the structure functions. These are explored in this paper via direct numerical simulations at various rotation rates.

The Kolmogorov K41 description of turbulence \cite{kolmogorov:local:1941,kolmogorov:refinement:1962} results in scaling laws for structure functions of the velocity increments. The second order structure function is the best-known, characterized by the famous Kolmogorov energy spectrum with a $-5/3$ slope. The Kolmogorov approach predicts a linear dependence of the scaling exponents on the order of the structure function~\cite{kolmogorov:local:1941}. However, in three-dimensional isotropic turbulence a so-called anomalous scaling of the structure functions is observed~\cite{kolmogorov:refinement:1962,sreenivasan:phenomenology:1997,biferale:isotropy:2001}. This is visible in a nonlinear dependence of the scaling exponents on the order of the structure function. This anomaly is associated with the effect of `intermittency'. One may expect that rotation, which induces a `trend' toward partial two-dimensionalization of the flow, will reduce intermittency and thereby also the anomalous scaling. Recent experiments in a freely decaying rotating turbulence using PIV show a strong increase of the exponents of the structure functions~\cite{seiwert:decrease:2008}. This is particularly pronounced for the second-order structure function. Correspondingly, a reduced scaling anomaly was reported. The main aim of this work is to complement these experimental findings with numerical simulations, allowing a direct correlation between reduced scaling anomaly and rotational flow structures. 

The organization of this paper is as follows. First, we introduce the computational setting for simulations of turbulence in a rotating frame of reference. Then, we present results of direct numerical simulations, quantifying the suppression of the energy decay for growing rotation rates. Afterwards, we quantify the intermittency via computation of the structure functions. The tendency to two-dimensionalize due to rotation is clearly expressed in a reduced presence of smaller scales in the flow and correspondingly increases scaling exponents. Finally, concluding remarks are collected in the last section.

\section{Computational setting} \label{sec:comp}

The decay of turbulence with an additional Coriolis force is investigated in a simple temporal setting using a parallelized, fully de-aliased pseudo-spectral method to simulate the flow in a computational box endowed with periodic boundary conditions. The incompressible Navier-Stokes equations for velocity $\mathbf{v}(\mathbf{x},t)$,  pressure $p(\mathbf{x},t)$, and external force $\mathbf{f}$:
\begin{equation}\label{eq:ns}
\left\{
\begin{array}{l}
\partial_t \mathbf{v} - \nu \Delta \mathbf{v} + (\mathbf{{v}} \cdot \nabla)\mathbf{v} + \nabla p = \mathbf{f}\\
\nabla \cdot \mathbf{v} =0\\
\end{array}
\right.
\end{equation}
are transformed into equations for the individual velocity Fourier coefficients $\mathbf{u}(\mathbf{k},t)$ in the wave-vector space $\mathbf{k}$:
\begin{equation}\label{eq:solve}
\Big( \partial_t + \nu k^2 \Big) \mathbf{u} - \mathbf{D}\mathbf{W} = \mathbf{F}.
\end{equation}
In the above equation, pressure is removed using the incompressibility condition via the projection operator $D_{\alpha \beta} = \delta_{\alpha \beta}-k_{\alpha} k_{\beta} /  k^2$; the nonlinear term $\mathbf{W} = -\mathcal{F} \big( (\mathbf{v} \cdot \nabla)\mathbf{v} \big)$ is transformed $\mathcal{F}(\cdot)$ to the Fourier space, but solved in the pseudo-spectral representation;  rotation is included via the external force also represented in the Fourier space as $\mathbf{F}= \mathcal{F}(\mathbf{f})$. Details of the code that is used for the numerical simulations can be found in~\cite{kuczaj:mixing:2006}. 

Rotation involves inclusion of the Coriolis and the centrifugal force. Assuming constant rotation rates, the centrifugal force can be incorporated in the pressure term and the Coriolis force reads $\mathbf{f} = -2 \mathbf{\Omega} \times \mathbf{v}$. In our simulations the rotation vector $\mathbf{\Omega}$ is assumed to be directed along the $z$-axis:  $\mathbf{\Omega } = \left[ {0, 0, \Omega} \right]$. Using the Levi-Civita permutation symbol $\epsilon$ and the Fourier representation, the external force for the rotation term can be finally written as: 
\begin{equation}
F_{\alpha} = -2 D_{\alpha \beta}\Omega \epsilon_{\beta 3 \gamma} {u_\gamma}
\end{equation}
In the numerical setup, the Navier-Stokes equations (\ref{eq:solve}) in the presence of constant rotation rate $\Omega$ along the $z$-axis are reformulated using the complex helical-wave decomposition framework~\cite{morinishi:new:2001}:
\begin{equation}\label{eq:hwd}
\Big[  \Big(  {\partial_t } + \nu k^2 \Big) \mathbf{I} + \frac{2\Omega k_3}{k^2} \mathbf{Q}^{-1} \mathbf{A} \mathbf{Q}  \Big] \mathbf{Q}^{-1}\mathbf{u} = \mathbf{Q}^{-1} \mathbf{DW},
\end{equation}
where $\mathbf{I}$ is the unit matrix. The matrix $\mathbf{A}$ is defined as:
\begin{equation}
\mathbf{A} = 
\left[ {\begin{array}{*{20}c}
{0} & {-k_3} & {+k_2} \\
{+k_3} & {0} & {-k_1} \\
{-k_2} & {+k_1} & {0} \\
\end{array}} \right],
\end{equation}
and the matrices $\mathbf{Q}$ and $\mathbf{Q}^{-1}$, corresponding to the eigenvalues of the matrix $\mathbf{A}$:  $\lambda_1 = +\imath k$,  $\lambda_2 = - \imath k$,  $\lambda_3 = 0 $ after solving the characteristic equation, have the form:
\begin{equation}
\mathbf{Q}^{~~} = q
\left[ {\begin{array}{*{20}c}
   s & s & {\sqrt {2k_1^2 s} }  \\
   { - k_1 k_2  - ikk_3 } & { - k_1 k_2  + ikk_3 } & {\sqrt {2k_2^2 s} }  \\
   { - k_1 k_3  + ikk_2 } & { - k_1 k_3  - ikk_2 } & {\sqrt {2k_3^2 s} }  \\
 \end{array} } \right],
\end{equation}
\begin{equation}
\mathbf{Q}^{ - 1}  = q
\left[ {\begin{array}{*{20}c}
   s & { - k_1 k_2  + ikk_3 } & { - k_1 k_3  - ikk_2 }  \\
   s & { - k_1 k_2  - ikk_3 } & { - k_1 k_3  + ikk_2 }  \\
   {\sqrt {2k_1^2 s} } & {\sqrt {2k_2^2 s} } & {\sqrt {2k_3^2 s} }  \\
 \end{array} } \right],
\end{equation}
where $q=({2k^2 s})^{-1/2} $ and $s={k_2^2  + k_3^2}$. The Coriolis force is diagonalized through this transformation and introduced as an integrating factor along with the viscous term in the Navier-Stokes equations. For time-integration the fourth-order, four-stage,  Runge-Kutta method is used. To remove the aliasing errors, the nonlinear term is computed on two grids (original and shifted) with an additional truncation to $m=120$ active modes for the lower resolution ($256^3$) and $m=241$ for the higher resolution ($512^3$) runs at the cut-off number $k_c = {2\pi} (m+1/2)/L_b$ for the cubic box of length $L_b=1$. Diagonalization of the Coriolis force through helical wave decomposition allows performing simulations at acceptable time-steps as the rotation does not dominate the reduction of the step in the explicit integration of the equations (\ref{eq:hwd}). Flow results obtained with direct numerical simulation are described next.

\begin{figure}[!hbt]
\begin{center}
\includegraphics[width=0.47\textwidth]{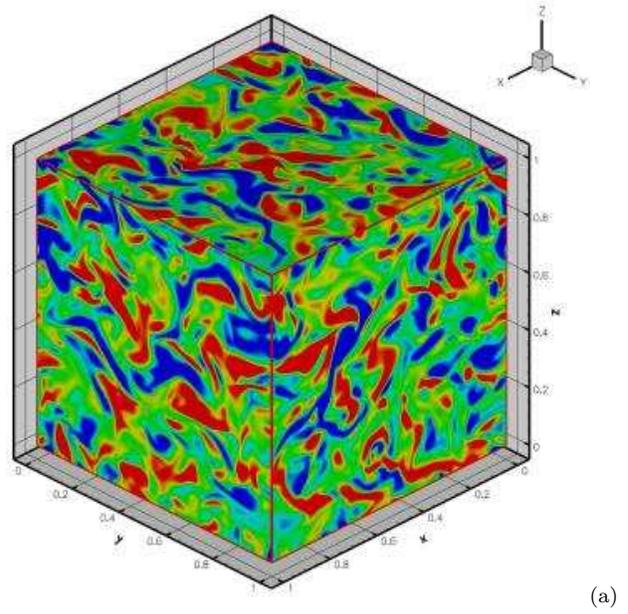}(a)
\includegraphics[width=0.47\textwidth]{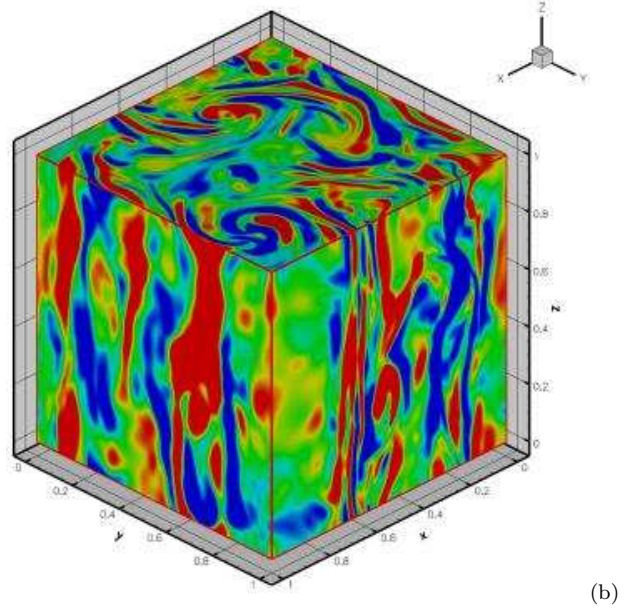}(b)
\end{center}
\caption{\label{fig:turb}Vertical structuring of rotating turbulence at $R_\lambda = 92$.
Snapshot of vorticity in $y$-direction at t = 2.5 for $Ro= \infty$~(a) and $Ro = 0.05$~(b).}
\end{figure}

\section{Direct Numerical Simulations}\label{sec:dns}

Direct numerical simulations of the spin-up of initially homogeneous isotropic turbulence at various rotation rates $\Omega$ and an initial Taylor-Reynolds numbers $R_\lambda= 50, 100, 200$ were performed. The decay of turbulence in the rotating setting at Rossby numbers $Ro=\frac{u_r}{2 \Omega L_b}= \infty, 0.5, 0.25, 0.1, 0.05, 0.025, 0.01, 0.005 $, which correspond to $\Omega=0, 1, 2, 5, 10, 20, 50, 100, 200$ at reference velocity $u_r=1$, is systematically investigated.  The initial condition for these simulations was taken from a simulation of non-rotating, homogeneous, isotropic decaying turbulence at a resolution of $512^3$. The generation of this condition is based on a random velocity field with prescribed spectrum, which is decayed for approximately two initial eddy-turnover times $\tau_0$: ($t=0.5 \approx 2\tau_0$, where $\tau_0$ denotes the initial eddy-turnover time). Subsequently, the evolved field is rescaled to the initial energy value ($E_0 = 0.5 u_r^2$) to serve as the initial condition in the simulations for various rotation rates. This procedure preserves the initial Reynolds number. Simulations at various Rossby numbers were performed with a resolution of $256^3$ for the lower Reynolds number and $512^3$ for higher turbulence levels. We simulated up to $t=6$ ($\approx 24 \tau_0$).

A snapshot showing the component of vorticity in the $y$-direction without rotation is presented in Fig.~\ref{fig:turb}(a). The small-scale vortical flow structures are seen to undergo striking qualitative changes when the system is rotating about the $z$-axis, see Fig.~\ref{fig:turb}(b) representing the flow at $Ro=0.05$. The rotational flow structuring corresponds to a trend toward (quasi) two-dimensional flow reminiscent of a strict Taylor-Proudman state. 

\begin{figure}[!hbt]
\begin{center}
\includegraphics[width=0.47\textwidth]{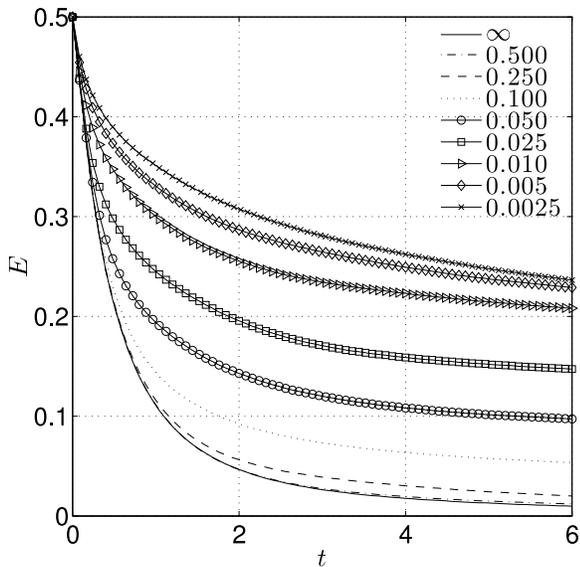}
\end{center}
\caption{\label{fig:deca} Decay of the energy $E(t)$ for various rotation rates (Rossby numbers) at the initial $R_{\lambda}=200$. Curves are labeled with $Ro$.}
\end{figure}

\begin{figure}[!hbt]
\begin{center}
\includegraphics[width=0.47\textwidth]{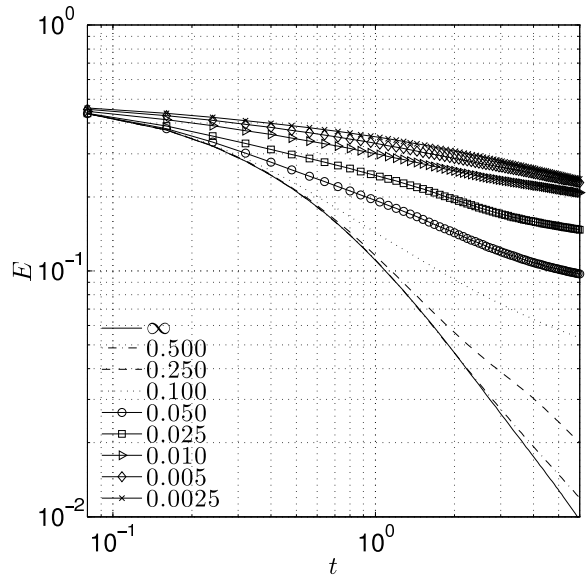}
\end{center}
\caption{\label{fig:decb} Log-log plot of the energy decay presented in Fig.~\ref{fig:deca}.}
\end{figure}

\begin{figure}[!hbt]
\begin{center}
\includegraphics[width=0.47\textwidth]{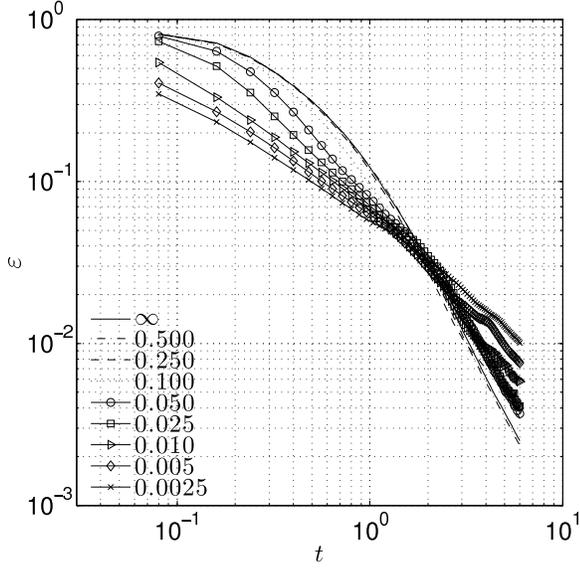}
\end{center}
\caption{\label{fig:decc} Energy dissipation rate $\varepsilon(t)$ for various rotation rates (Rossby numbers) at the initial $R_{\lambda}=200$.}
\end{figure}

Corresponding to the much larger coherent structures in the flow, the decay rate of the kinetic energy $E$ is strongly reduced due to rotation as may be observed in Fig.~\ref{fig:deca}.  The energy decay is usually characterized by a decay exponent $n$ such that $E(t) \sim t^{n}$ for large times~\cite{speziale:energy:1991,george:decay:1992}. In case of self-similar decay consistent with  $E(t) \sim t^{n}$, the decay exponent $n$ should be constant in time. Hence, the results presented in Fig.~\ref{fig:deca} should form straight lines when shown on a log-log scale. The decay exponent associated with the slope of the decay can then be easily examined. In our simulations, two separate regions can be associated with the decay of the energy. Only for sufficiently large times (see Fig.~\ref{fig:decb}) do we observe self-similarity. In the first period up to about $t=1$, the energy decay slope significantly varies in the first period of the decay. This is in particular visible for low rotation rates. Such behavior can be understood from the fact that at low rotation rates the flow needs more time to adapt itself to the imposed rotation. The isotropic homogeneous turbulence is preserved for a longer time in such cases. For high rotation rates, all flow scales are strongly influenced and a transitional period before reaching the self-similar stages is much less pronounced.  Consequently, the self-similar decay of energy can be safely examined for $t>1$. The structure functions were also computed for $t>1$. The decay exponent is observed to decrease as the rotation rate is increased. Such behavior of the decay exponent is in agreement with results obtained for the experimentally investigated grid-generated turbulence in a rotating tank~\cite{morize:energy:2006}. During the initial stage of the decay, the energy decay is more suppressed for higher rotation rates and consequently the energy dissipation rate has much smaller values in the initial phase of the decay, see Fig.~\ref{fig:decc}. Similarly, as in case of the kinetic energy, the self-similar decay of the energy dissipation rate is present for large $t$ only. This is particularly true for the smaller rotation rates.

\begin{figure}[!hbt]
\begin{center}
\includegraphics[width=0.47\textwidth]{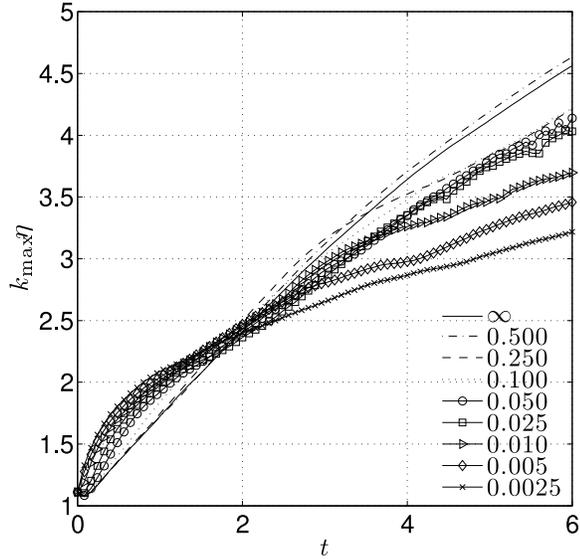}
\end{center}
\caption{\label{fig:km} Resolution $k_{\max} \eta$ criterion for various Rossby numbers as a function of time at the initial $R_{\lambda}=200$.}
\end{figure}

To assess the resolution of the flow we examine the so-called $k_{\max}\eta$ criterion, where $\eta$ is the Kolmogorov length and $k_{\max}$ is the largest wavenumber used in the simulations. The result is shown in Fig.~\ref{fig:km}. In order to resolve dynamically relevant turbulence length-scales, it is required that $k_{\max} \eta$ is sufficiently large. A commonly accepted criterion of adequate spatial resolution is that $k_{\max}\eta > 1$. For all rotation rates, this criterion is satisfied. This underpins confidence that the Kolmogorov scale is properly resolved in our simulations.
Larger anisotropic structures built by backward energy transfer for larger rotation rates.  As the flow develops in time, the energy decays further and the energy dissipation rate decreases to very small values. Subsequently, quite large values of the Kolmogorov scale $\eta$ can be observed for the final periods of the decay.

\begin{figure}[!hbt]
\begin{center}
\includegraphics[width=0.47\textwidth]{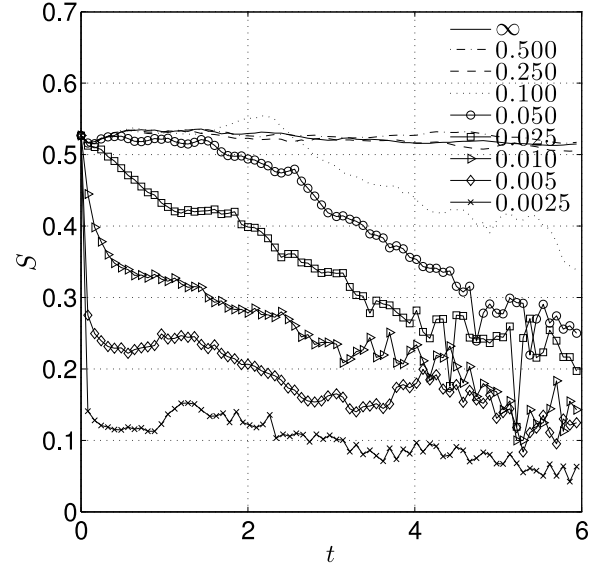}
\end{center}
\caption{\label{fig:S} Skewness $S$ for various rotation rates (Rossby numbers) as a function of time at the initial $R_{\lambda}=200$.}
\end{figure}

Rotation changes the statistical properties of the turbulent flow. For example skewness is reduced from its theoretical value of $S=0.5$ in homogeneous isotropic turbulence as proven by Batchelor \cite{batchelor:theory:1953}, cf. Fig.~\ref{fig:S}. In our simulations we use the spectral formula for the computation of the skewness: 
\begin{equation}
S(t) = \frac{2} {{35}}\left( {\frac{{\lambda (t)}} {{u(t)}}} \right)^3 \int\limits_0^{k_{\max } } {k^2 T(k,t)dk},
\end{equation}
where $\lambda(t)$ is the Taylor microscale, $u(t)$ is the rms velocity, and $T(k,t)$ is the energy transfer term at time $t$ and wavenumber $k$. The observed significant modulation of skewness for higher rotation rates can also be associated   
with a departure from isotropy in the considered flow. The current simulations display an immediate acceleration in the initial period of the flow development. This immediate change in the flow properties for high rotation rates is seen through a significant reduction of skewness in the first time-steps of the simulations. These initial data are not taken into consideration in our analyses presented here. 

Next we turn our attention to the intermittency effects, which recently were measured in the same experimental facility~\cite{seiwert:decrease:2008} as the decay of turbulence examined earlier~\cite{morize:energy:2006}.

\section{Intermittency} \label{sec:intermit}

The structure functions used in our setup have the following form:
\begin{equation}
S_p (r) = \left\langle {\left| {\delta v_r } \right|^p } \right\rangle,
\end{equation}
where $p$ is the order of the structure function and $\delta v_r = ( \mathbf{v} (\mathbf{x}+\mathbf{r}) - \mathbf{v}) \cdot \mathbf{r}/r$ is the longitudinal velocity increment computed as the volume average $\left\langle \cdot \right\rangle$. The separation distance $\mathbf{r}$ normal to the rotation vector $\mathbf{\Omega}$. Intermittency usually refers to the anomalous scaling $\varsigma _p$  in the similarity hypothesis applied for the structure functions:
\begin{equation}
S_p (r)\propto r^{\varsigma _p },
\end{equation}
The scaling exponents were computed from a numerical derivative: $\varsigma _p  = ({d\log(S_p (r))}) / ({d\log{r}})$.

Theory of homogeneous isotropic turbulence predicts that in a self-similar flow an inertial range exists in which $\varsigma_p = p/3$~\cite{frisch:legacy:1995} and correspondingly, e.g., the energy spectrum is described as $E(k) \sim k^{-5/3}$. In contrast, for rotating turbulence much steeper energy spectra are predicted by Kraichnan assuming a totally inhibited energy transfer. This was predicted to yield $E(k)\sim k^{-3}$ or $E(k)\sim k^{-2}$ for a  spectrum dictated by an energy transfer time-scale proportional to $1/\Omega$~\cite{zhou:phenomenological:1995}. Subsequently, the anomalous scaling of the structure function $\varsigma_p =p/2$ was  experimentally found in rapidly rotating fluids~\cite{baroud:anomalous:2002}. Reduced intermittency measured as a departure from the $\varsigma_p/\varsigma_2 =p/2$ will be shown here for the rotating decaying turbulence.

\begin{figure}[!hbt]
\begin{center}
\includegraphics[width=0.47\textwidth]{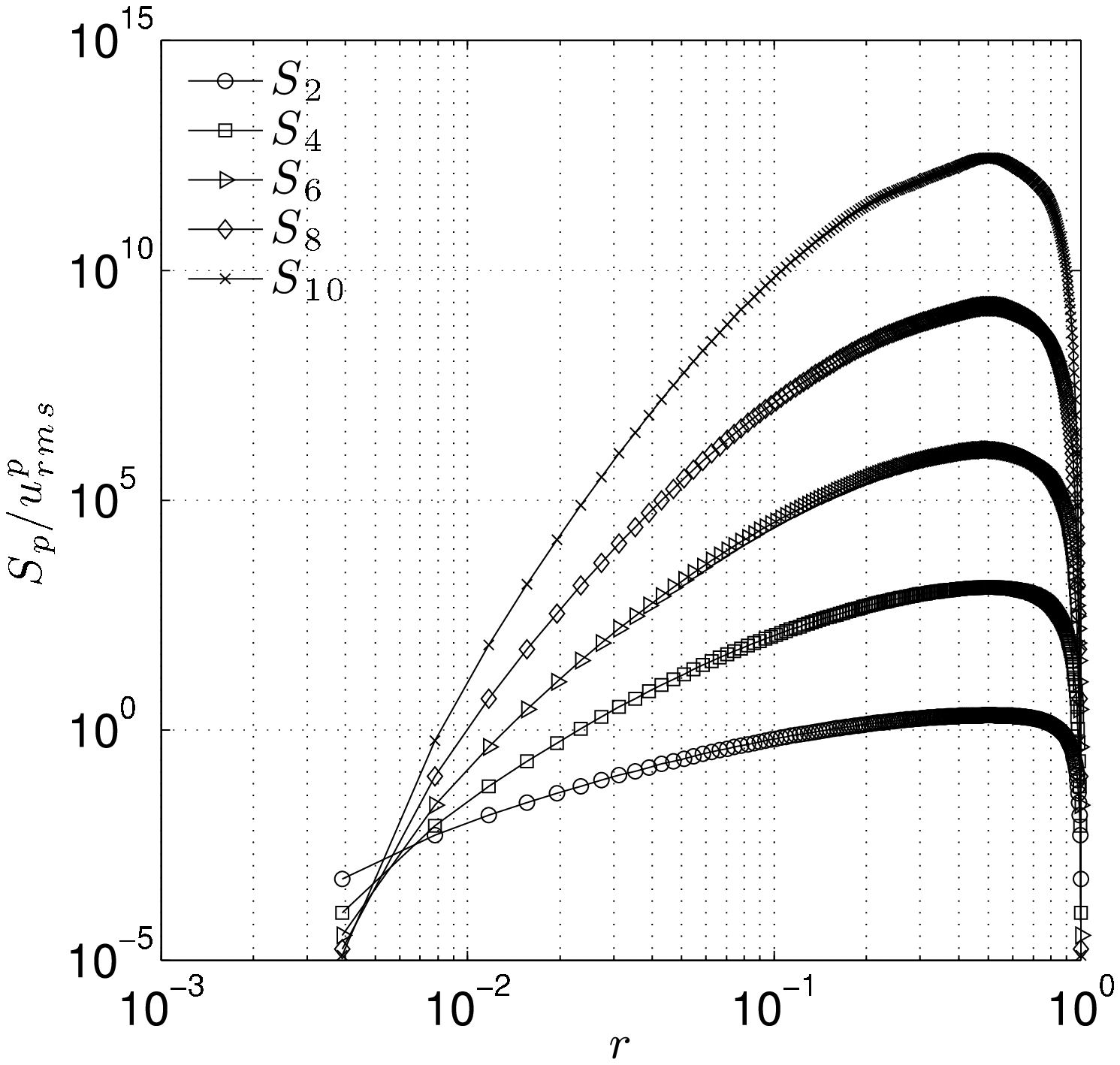}(a)
\includegraphics[width=0.47\textwidth]{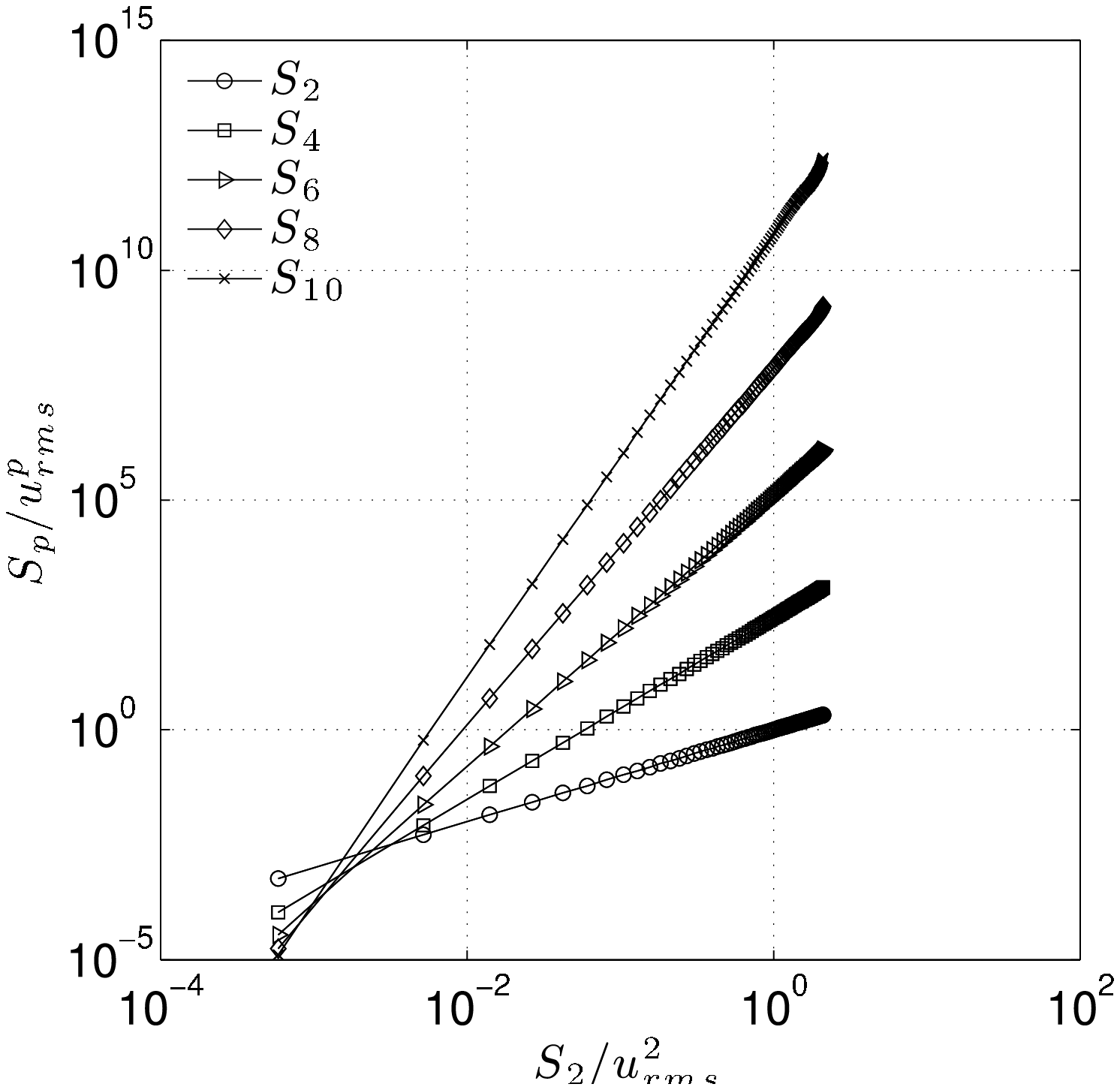}(b)
\caption{\label{fig:1} Structure functions for increasing order at an initial $R_\lambda =200$ at $t=2$ and rotation rate $\Omega=1$ ($Ro=0.5$) plotted as a function of the separation scale $r$ (a) and as a function of $S_2$~(b). The curves for $p=4,6,8,10$ have been vertically shifted by factors $10^2, 10^4, 10^6$ and $10^8$ for visibility.}
\end{center}
\end{figure}

An example of the structure functions for increasing order $p$ is shown in Fig.~\ref{fig:1} (compare with Fig.~2 in \cite{seiwert:decrease:2008}). The separations $\mathbf{r}$ are normal 
to the rotation vector $\Omega$. This is particularly important, as the observed departures from the isotropic properties of turbulence are analyzed in a direction, which is perpendicular to the rotation axis.  Hence, they are not directly influenced by the anisotropic character of the flow in this direction. The presented investigations are restricted to the self-similar range $t>1$ and compared to the ESS~\cite{benzi:extended:1993,grossmann:application:1997} method.

\begin{figure}[!hbt]
\begin{center}
\includegraphics[width=0.47\textwidth]{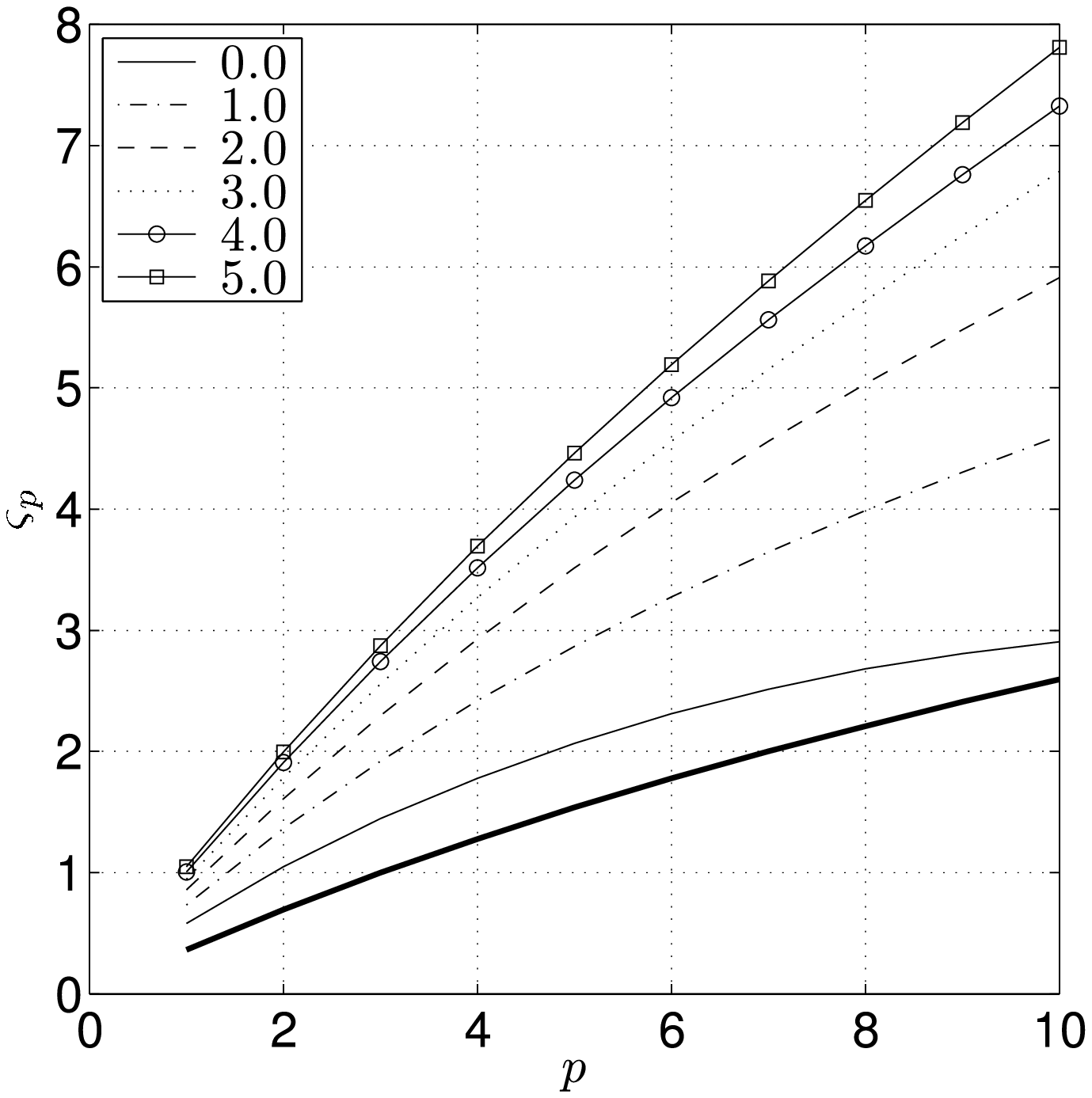}(a)
\includegraphics[width=0.47\textwidth]{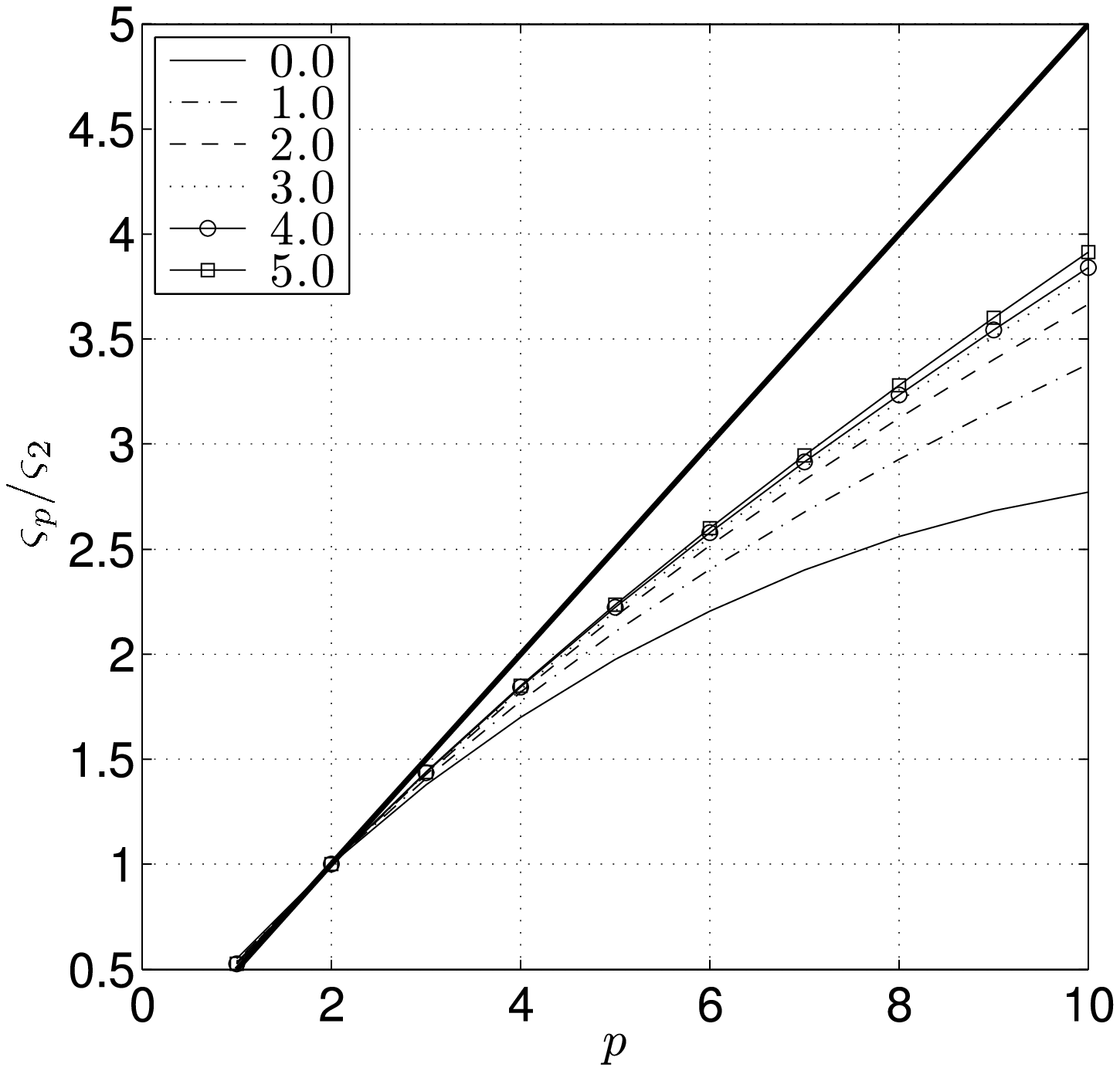}(b)
\caption{\label{fig:3} Structure functions exponents  (a) and normalized exponents $\varsigma_p / \varsigma_2$ at an  initial $R_\lambda =200$ at various times during the decay used to label the curves, using  $\Omega=1$. Scaling exponents obtained from She-Leveque (a) and $p/2$  (b) are indicated with the thick line.}
\end{center}
\end{figure}

Our DNS results show modulation of the structure functions due to rotation, as illustrated  in Fig.~\ref{fig:1} for a particular instance in time and low rotation rate ($Ro=0.5$). The raw $\varsigma_p$ and normalized exponents $\varsigma_p / \varsigma_2$ at different times during the decay are shown in Fig.~\ref{fig:3}. Similarly, as in the experimental results~\cite{seiwert:decrease:2008}, the exponents differ from the strictly linear $p/2$ law reported in~\cite{baroud:anomalous:2002}.  

\begin{figure}[!hbt]
\begin{center}
\includegraphics[width=0.47\textwidth]{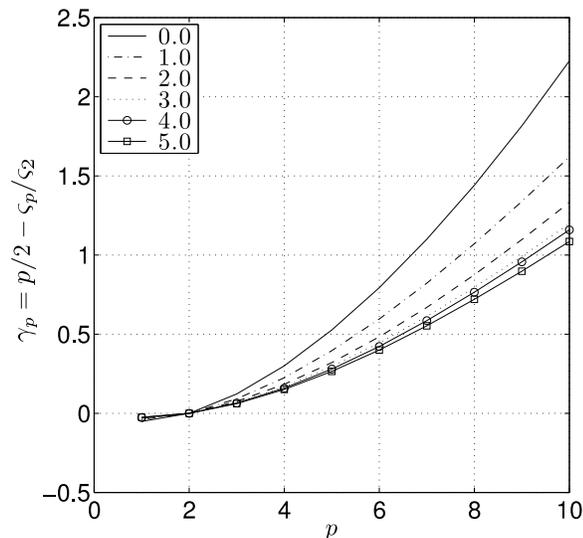}
\caption{\label{fig:4} Intermittency factors  $\gamma_p = p/2 -\varsigma_p/\varsigma_2$ for large Rossby number $Ro=0.5$.}
\end{center}
\end{figure}

The reduction of intermittency can be clearly seen from the intermittency factors  $\gamma_p = p/2 -\varsigma_p/\varsigma_2$, which should vanish for non-intermittent velocity fluctuations. These factors were plotted for various instances in time and low rotation rate in Fig.~\ref{fig:4}. They show similar behavior as in the experiment with a significant decrease during the initial stages of the flow development. This behavior is probably connected to the initial rapid spin-up of the flow that causes immediate two-dimensionalization of the flow, i.e., the formation of elongated flow structures along the axis of rotation.

\section{Summary}\label{sec:sum}
Our numerical simulations of rotating decaying turbulence indicate a strong increase of the structure function exponents during the decay. The observed modulation of the structure functions is directly associated with changes in the flow characteristics, i.e., the increase of the structure function exponents is related to the much steeper energy spectra in case of rotation. An important fact is that the structure functions are measured in a direction perpendicular to the axis of rotation. This implies that the observed changes of the turbulence characteristics appear in a direction, which is not directly influenced by the anisotropy introduced through the Coriolis force. Non-negligible deviations from Kolmogorov-universality and self-similarity are observed as a result of anisotropization of the flow. This departure can be directly linked with the modulation/interplay of the intermittency phenomenon and scale-dependent energy transfer in the energy cascade process (deviation from homogeneity). As a result, it may be inferred that in case of rotating turbulence, both the longitudinal and transverse components of the structure functions are equally relevant, as the contribution to the energy transfer from the latter component does not statistically vanish because of homogeneity as in the K41/K62 framework. In particular, the presented results for rotating decaying flow are in line with experimental findings obtained in a rotating tank with background turbulence~\cite{seiwert:decrease:2008} and forced flow conditions obtained through direct numerical simulations~\cite{muller:scaling:2007}. Strong increase of the second order scaling exponents suggests that in case of rotation, steeper energy spectra are predicted than those obtained under the assumption of nonlinear interactions leading to an energy transfer governed by the time-scale $1/\Omega$: $E(k) \sim k^{-2}$~\cite{zhou:phenomenological:1995}. Further research will be devoted to this subject, in which energy spectra exponents will be analyzed.

\section{Acknowledgments}
AKK is grateful to Frederic Moisy for a fruitful discussion at the Isaac Newton Institute for Mathematical Sciences in Cambridge. Computations were performed at the~SARA Computing and Networking Services in Amsterdam, which were made
possible through grant of the Dutch National Computing Foundation (NCF).

\bibliographystyle{unsrt}

\begin{thebibliography}{10}

\bibitem{seiwert:decrease:2008}
J.~Seiwert, C.~Morize, and F.~Moisy.
\newblock On the decrease of intermittency in decaying rotating turbulence.
\newblock {\em Phys. Fluids}, 20(071702), 2008.

\bibitem{chen:resonant:2005}
Q.~Chen, S.~Chen, G.L. Eyink, and D.D. Holm.
\newblock Resonant interactions in rotating homogeneous three-dimensional
  turbulence.
\newblock {\em J. Fluid Mech.}, 542:139--164, 1997.

\bibitem{kolmogorov:local:1941}
A.N. Kolmogorov.
\newblock The local structure of turbulence in incompressible viscous fluids at
  very large {Reynolds} numbers.
\newblock {\em C.R. Acad. Sci. URSS}, 30:301--305, 1941.

\bibitem{kolmogorov:refinement:1962}
A.N. Kolmogorov.
\newblock A refinement of previous hypothesis concerning the local structure of
  turbulence in a viscous incompressible fluid at high {Reynolds} number.
\newblock {\em J. Fluid Mech.}, 13:82--85, 1962.

\bibitem{sreenivasan:phenomenology:1997}
K.R. Sreenivasan and R.A. Antonia.
\newblock The phenomenology of small-scale turbulence.
\newblock {\em Ann. Rev. Fluid Mech.}, 29:435--472, 1997.

\bibitem{biferale:isotropy:2001}
L.~Biferale and M.~Vergassola.
\newblock Isotropy vs anisotropy in small-scale turbulence.
\newblock {\em Phys. Fluids}, 13:2139, 2001.

\bibitem{kuczaj:mixing:2006}
A.K. Kuczaj and B.J. Geurts.
\newblock Mixing in manipulated turbulence.
\newblock {\em J. Turbul.}, 7 (N67), 2006.

\bibitem{morinishi:new:2001}
Y.~Morinishi, K.~Nakabayashi, and S.Q. Ren.
\newblock A new {DNS} algorithm for rotating homogeneous decaying turbulence.
\newblock {\em Int. J. of Heat and Fluid Flow}, 22:30--38, 2001.

\bibitem{speziale:energy:1991}
Ch.G. Speziale and P.S. Bernard.
\newblock The energy decay in self-preserving isotropic turbulence revisited.
\newblock {\em NASA ICASE Report}, 91(58), 1991.

\bibitem{george:decay:1992}
W.K. George.
\newblock The decay of homogeneous isotropic turbulence.
\newblock {\em Phys. Fluids A}, 4(7):1492--1509, 1992.

\bibitem{morize:energy:2006}
C.~Morize and F.~Moisy.
\newblock Energy decay of rotating turbulence with confinement effects.
\newblock {\em Phys. Fluids}, 18(065107), 2006.

\bibitem{batchelor:theory:1953}
G.~K. Batchelor.
\newblock {\em Theory of homogeneous turbulence}.
\newblock Cambridge University Press, 1953.

\bibitem{frisch:legacy:1995}
U.~Frisch.
\newblock {\em Turbulence, the legacy of A.N. Kolmogorov}.
\newblock Cambridge University Press, 1995.

\bibitem{zhou:phenomenological:1995}
Y.~Zhou.
\newblock A phenomenological treatment of rotating turbulence.
\newblock {\em Phys. Fluids}, 7:2092, 1995.

\bibitem{baroud:anomalous:2002}
Ch.N. Baroud, B.B. Plapp, Z.-S. She, and H.L. Swinney.
\newblock Anomalous self-similarity in a turbulent rapidly rotating fluid.
\newblock {\em Phys. Rev. Lett.}, 88(11):114501, 2001.

\bibitem{benzi:extended:1993}
R.~Benzi, S.~Ciliberto, R.~Tripiccione, C.~Baudet, F.~Massaioli, and S.~Succi.
\newblock {Extended Self Similarity} in turbulent flow.
\newblock {\em Phys. Rev. E}, 48(R29), 1993.

\bibitem{grossmann:application:1997}
S.~Grossmann, D.~Lohse, and A.~Reeh.
\newblock Application of extended self-similarity in turbulence.
\newblock {\em Phys. Rev. E}, 56:5473--5478, 1997.

\bibitem{muller:scaling:2007}
W.-C. Muller and M.~Thiele.
\newblock Scaling and energy transfer in rotating turbulence.
\newblock {\em EPL}, 77 (34003):5, 2007.

\end{thebibliography}

\end{document}